\documentclass[%a
 reprint,
superscriptaddress,
 amsmath,amssymb,
 aps, physrev,
]{revtex4-2}
\usepackage[version=3]{mhchem} % Formula subscripts using \ce{}
\usepackage{float}
\usepackage{comment}
\usepackage{esint}
\usepackage{tabularx}
\usepackage{soul}
\usepackage{amsmath}
\usepackage{amssymb}
\usepackage[normalem]{ulem}

\usepackage[utf8]{inputenc}
\usepackage[T2A,T1]{fontenc}
\usepackage[english]{babel}
\usepackage[
colorlinks=true,bookmarks=false,citecolor=blue,urlcolor=blue
]
{hyperref}

\usepackage{graphicx}% Include figure files
\usepackage{color}
\usepackage{soul}
\usepackage{xcolor}
\usepackage{comment}

\usepackage{xfrac}
\usepackage{gensymb}
\usepackage[caption=false]{subfig}

\usepackage{dcolumn}% Align table columns on decimal point
\usepackage{bm}% bold math

\definecolor{gr}{RGB}{86,184,57}

\begin{document}
\preprint{APS/123-QED}
\title{\textbf{Broadband wide-view all-dielectric handedness-preserving mirror}} 
%%%%%%%%%%%%%%%%%%%%%%%%%%%%%%%%%%%%%%%%%%%%%%%%%%%%%%%%%%%%%%%%%%%%%%%%%%%%%%%%%%%%%%%%%%%%%%%%%%%%%%%%%%%%%%%%%%
\author{Natalia Salakhova}
\affiliation{Skolkovo Institute of Science and Technology, Moscow, Russia}

\author{Andrey Demenev}
\affiliation{Osipyan Institute of Solid State Physics RAS, Chernogolovka, Russia}

\author{Oleg Klimenko}
\affiliation{Skolkovo Institute of Science and Technology, Moscow, Russia}
\affiliation{PN Lebedev Physical Institute of RAS, Leninskiy prospekt 53, Moscow, Russia 119991}

\author{Vladimir Kulakovskii}
\affiliation{Osipyan Institute of Solid State Physics RAS, Chernogolovka, Russia}

\author{Vladimir Antonov}
\affiliation{Skolkovo Institute of Science and Technology, Moscow, Russia}

\author{Nikolay Gippius}
\affiliation{Skolkovo Institute of Science and Technology, Moscow, Russia}

\author{Sergey Dyakov}
\affiliation{Skolkovo Institute of Science and Technology, Moscow, Russia}

\date{\today}
%%%%%%%%%%%%%%%%%%%%%%%%%%%%%%%%%%%%%%%%%%%%%%%%%%%%%%%%%%%%%%%%%%%%%%%%%%%%%%%%%%%%%%%%%%%%%%%%%%%%%%%%%%%%%%%%%%
\begin{abstract}
We report the theoretical design and experimental realization of a wideband, all-dielectric mirror that preserves the handedness of incident light upon reflection in the near-infrared range. The mirror consists of a high-contrast, near-subwavelength, one-dimensional dielectric grating on a Bragg mirror. We optimized this structure using a genetic algorithm and demonstrated its robustness against geometric imperfections and oblique incidence. Experimental reflection spectra measured under normal incidence in a circular polarization basis demonstrate a more than 100-nm-wide reflection band in which more than 98\% of the reflected light preserves its handedness. The total reflection coefficient reached 80\%. Furthermore, we demonstrate that the fabricated mirror maintains high performance even under oblique incidence for angles up to $\pm 15^\circ$. Owing to these unique characteristics, this mirror can serve as a reflective phase plate, making it an excellent candidate for the creation of a Fabry–Pérot resonator for chiral light.
\end{abstract}

\maketitle
\newpage

%%%%%%%%%%%%%%%%%%%%%%%%%%%%%%%%%%%%%%%%%%%%%%%%%%%%%%%%%%%%%%%%%%%%%%%%%%%%%%%%%%%%%%%%%%%%%%%%%%%%%%%%%%%%%%%%%%

\section*{Introduction}

Light-matter interactions involving optical chirality have attracted significant attention over the past decade, driven by the rapid development of chiral metasurfaces, polarization-selective photonic devices, and compact platforms for controlling the handedness of light. A fundamental limitation of conventional homogeneous isotropic mirrors is that they invert the handedness of circularly polarized light upon reflection. Overcoming this limitation and achieving handedness-preserving reflection is crucial for various fields of applied photonics. 

One of the most intriguing applications related to optical chirality is the development of highly sensitive sensors for chiral molecules. This challenge is essential in pharmaceutical and biological systems, where the specific chirality of a compound can determine its bioactivity and toxicity \cite{kasprzyk2010pharmacologically,ceramella2022look}. A promising route toward enhanced enantioselective detection involves nanophotonic structures that resonantly amplify the weak interaction between chiral light and chiral matter \cite{Tang2011,warning2021nanophotonic,Kumar2024,dyakov2025strong}. Resonators supporting high-quality chiral electromagnetic modes can interact differently with left- and right-handed molecules, allowing highly sensitive measurements from small analyte volumes \cite{zhao2017chirality,zhang2024quantum}. Various platforms, including chiral metamaterials \cite{Kan2015,FernandezCorbaton2019,zhao2017chirality,deng2024advances}, plasmonic helices \cite{Hendry2010,Pham2016,Barbillon2020,Wang2023}, nanoparticles \cite{Wu2013,Graf2019}, and complex chiral metasurfaces \cite{petersen2014chiral,shi2022planar,konishi2011circularly,Mohammadi2019,gorkunov2020metasurfaces}, have been explored for this purpose. Among these, the Fabry-Pérot architecture \cite{feis2020helicity,Voronin2022,gautier2022planar,dyakov2024chiral} remains particularly attractive because of its conceptual simplicity, strong field confinement, and compatibility with established fabrication processes.

As long as conventional homogeneous isotropic mirrors invert the handedness of light upon reflection, the key requirement for implementing a Fabry-Pérot resonator supporting chiral modes is the creation of a handedness-preserving mirror \cite{Voronin2022,dyakov2024chiral}. Several approaches to the implementation of handedness-preserving reflection have been proposed, including three-dimensional chiral metamaterials \cite{ma2017meta,plum2016extrinsic,kang2017preserving}, anisotropic multilayer coatings \cite{rudakova2018all}, and low-symmetric periodic metasurfaces \cite{Plum2015,Semnani2020,li2020spin,Voronin2022,dyakov2024chiral,fradkin2023nearly}. Ref.~\cite{Semnani2020,Voronin2022} reported in-plane chiral metasurfaces that demonstrate near-perfect reflection of one circular polarization with handedness preservation and transparency for the opposite polarization. It has been shown that the required functionality is achieved when the structure has a pair of orthogonal eigenstates with opposite parity \cite{Voronin2022}.

The resonant nature of low-symmetric handedness-preserving mirrors significantly constricts their operating bandwidth. This fact, along with fabrication complexity, substantially limits the practical applicability of resonant metasurfaces.

In this work, we demonstrate a non-resonant broadband handedness-preserving mirror based on a one-dimensional dielectric grating placed on top of a distributed Bragg reflector. The grating introduces control of the in-plane anisotropy, while the Bragg mirror ensures high reflectivity and stabilizes the spectral phase response. We show that together, these elements provide handedness preservation upon reflection across a wide wavelength range. The operating range of the proposed metasurface lies within the stop-band of the Bragg mirror. The design relies solely on standard dielectric materials and fully planar fabrication, making it compatible with scalable nanolithography.

%%%%%%%%%%%%%%%%%%%%%%%%%%%%%%%%%%%%%%%%%%%%%%%%%%%%%%%%%%%%%%%%%%%%%%%%%%%%%%%%%%
%%%%%%%%%%%%%%%%%%%%%%%%%%%%%%%%%%%%%%%%%%%%%%%%%%%%%%%%%%%%%%%%%%%%%%%%%%%%%%%%%%
%%%%%%%%%%%%%%%%%%%%%%%%%%%%%%%%%%%%%%%%%%%%%%%%%%%%%%%%%%%%%%%%%%%%%%%%%%%%%%%%%%
\begin{figure*}[t!]
\centering\includegraphics[width=1\textwidth]{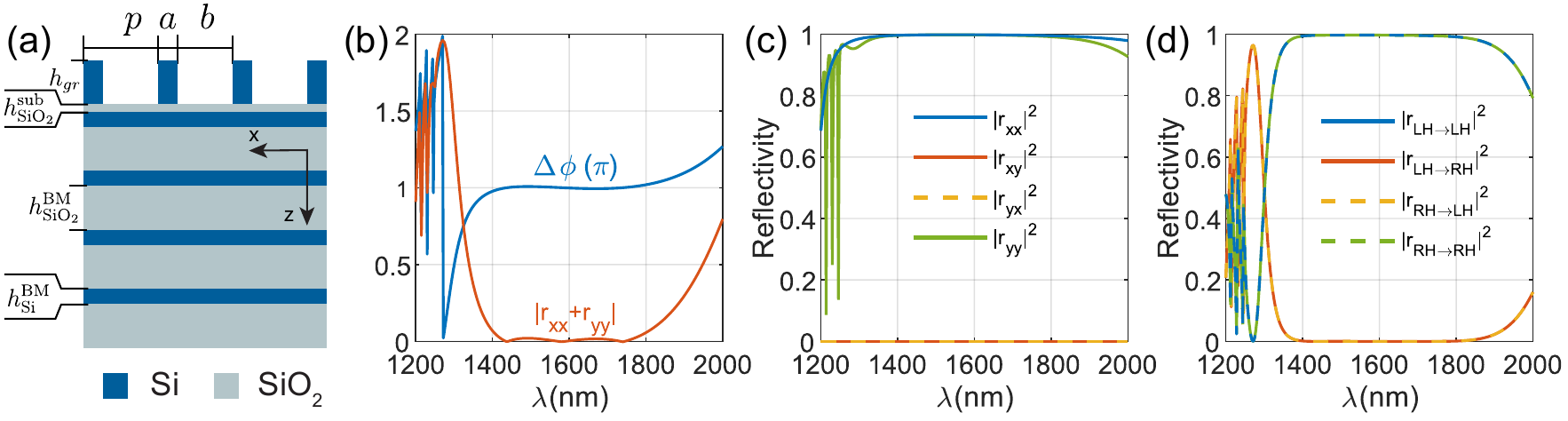}
\caption{Modeling. (a) The sketch of a wide-band handedness-preserving mirror consisting of the one-dimensional Si grating on a Si/SiO$_2$ multilayered structure. (b) Spectral dependence of $\Delta \phi = \arg(r_{xx})-\arg(r_{yy})$ and $|r_{xx}+r_{yy}|$. (c) Spectral dependencies of cross-polarization reflection coefficients in the basis of linear polarizations. (d) Spectral dependencies of cross-polarization reflection coefficients in the basis of circular polarizations. Parameters used for modeling: thicknesses of the layers was $h_{gr} = 297$~nm (grating), $h_{\text{SiO}_2}^{\text{sub}} = 50$~nm (SiO$_2$ sublayer), $h_{\text{Si}}^{\text{BM}} = 102$ nm (Si layer of the Bragg mirror), $h_{\text{SiO}_2}^{\text{BM}} = 297$~nm (SiO$_2$ layer of the Bragg mirror); the period $p = 503$~nm; the stripes' width $a = 131$~nm the grooves' width $b = 372$~nm.}
\label{fig:mirror}
\end{figure*}

\begin{figure}[t!]
\centering\includegraphics[width=1\linewidth]{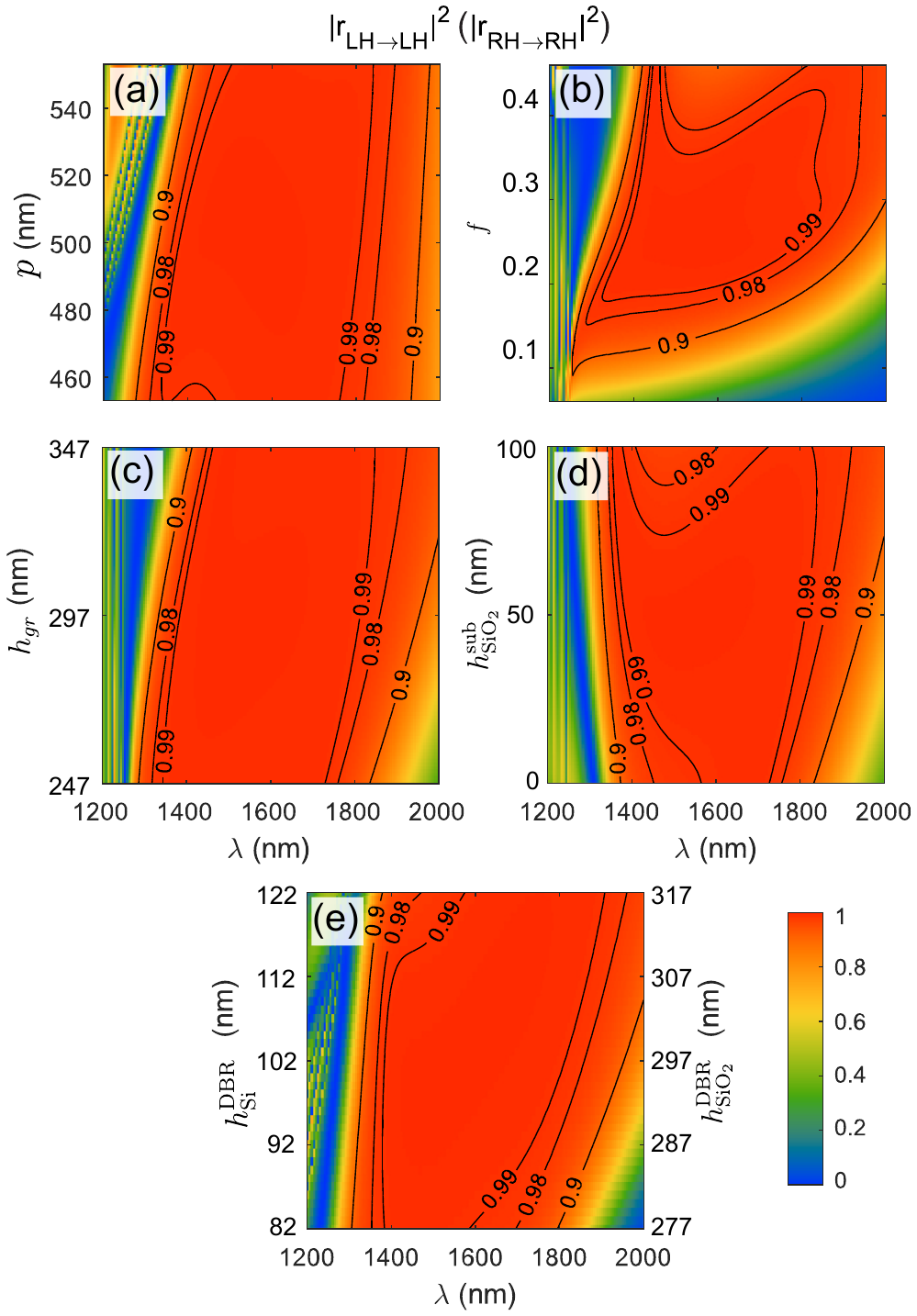}
\caption{Modeling. Reflection coefficient for LH→LH (RH→RH) polarization as a function of the light wavelength and structural parameters: (a) grating period $p$, (b) filling factor $f = a/p$, (c) grating thickness $h_{gr}$, (d) SiO$_2$ sublayer thickness $h_{\text{SiO}_2}^{\text{sub}}$, and (e) Bragg mirror layers thickness $h_{\text{Si}}^{\text{BM}}$, $h_{\text{SiO}_2}^{\text{BM}}$ that was changed simultaneously. Initial parameters used for modeling: $h_{gr} = 297$~nm, $h_{\text{SiO}_2}^{\text{sub}} = 50$~nm, $h_{\text{Si}}^{\text{BM}} = 102$ nm, $h_{\text{SiO}_2}^{\text{BM}} = 297$~nm, $p = 503$~nm, $a = 131$~nm, $b = 372$~nm.}
\label{fig:geom}
\end{figure}

\begin{figure}[t!]
\centering\includegraphics[width=1\linewidth]{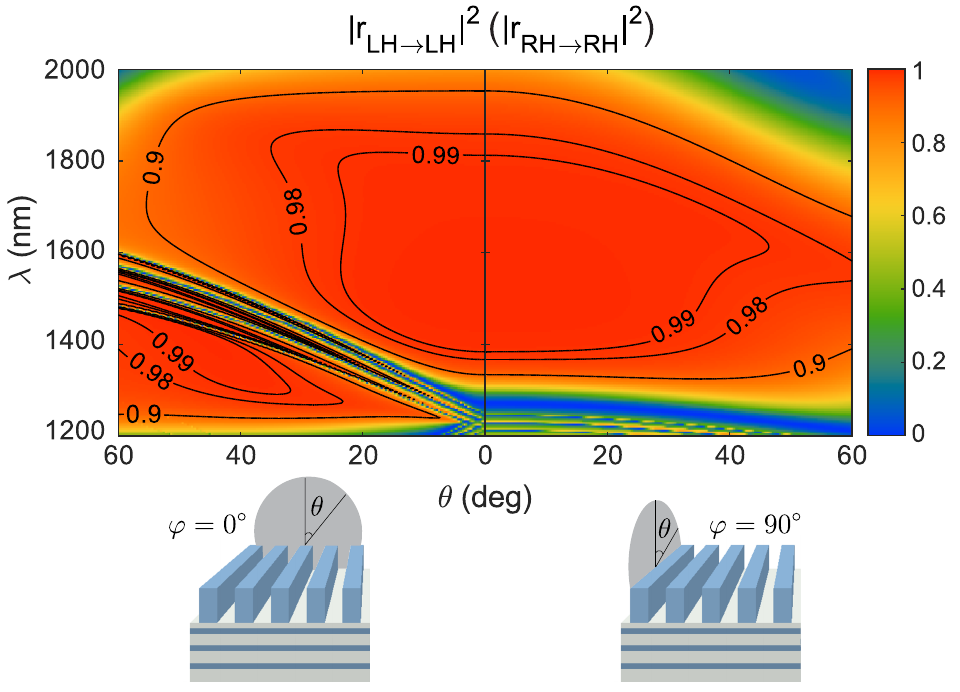}    
\caption{Modeling. Reflection coefficient for LH→LH (RH→RH) polarization as a function of the light wavelength and incidence angle $\theta$: incidence perpendicular to the grating stripes and parallel to the grating stripes. Parameters used for modeling: $h_{gr} = 297$~nm, $h_{\text{SiO}_2}^{\text{sub}} = 50$~nm, $h_{\text{Si}}^{\text{BM}} = 102$ nm, $h_{\text{SiO}_2}^{\text{BM}} = 297$~nm, $p = 503$~nm, $a = 131$~nm, $b = 372$~nm.}
\label{fig:angle}
\end{figure}

%\begin{figure*}[t!]
%\centering\includegraphics[width=0.9\linewidth]{Fig4_2_eff_3.pdf}    %UP3.pdf
%\caption{Experiment. (a) SEM image of the grating on the top of the Bragg mirror. Reflection spectra and co-handedness reflection efficiency for (b)-(c) normal incidence, (d)-(e) $\theta = 5^{\circ}$ and $\varphi = 0^{\circ}$ and (f)-(g) $\theta = 5^{\circ}$ and $\varphi = 9 0^{\circ}$. Approximate parameters of fabricated structure: $h_{gr} =  290$~nm, $h_{\text{SiO}_2}^{\text{sub}} = 60$~nm, $h_{\text{Si}}^{\text{BM}} = 110$~nm, $h_{\text{SiO}_2}^{\text{BM}} = 260$~nm, $p = 650$~nm, $a = 150$~nm, $b = 500$~nm.}
%\label{fig:exp1}
%\end{figure*}

\begin{figure}[t!]
\centering\includegraphics[width=1\linewidth]{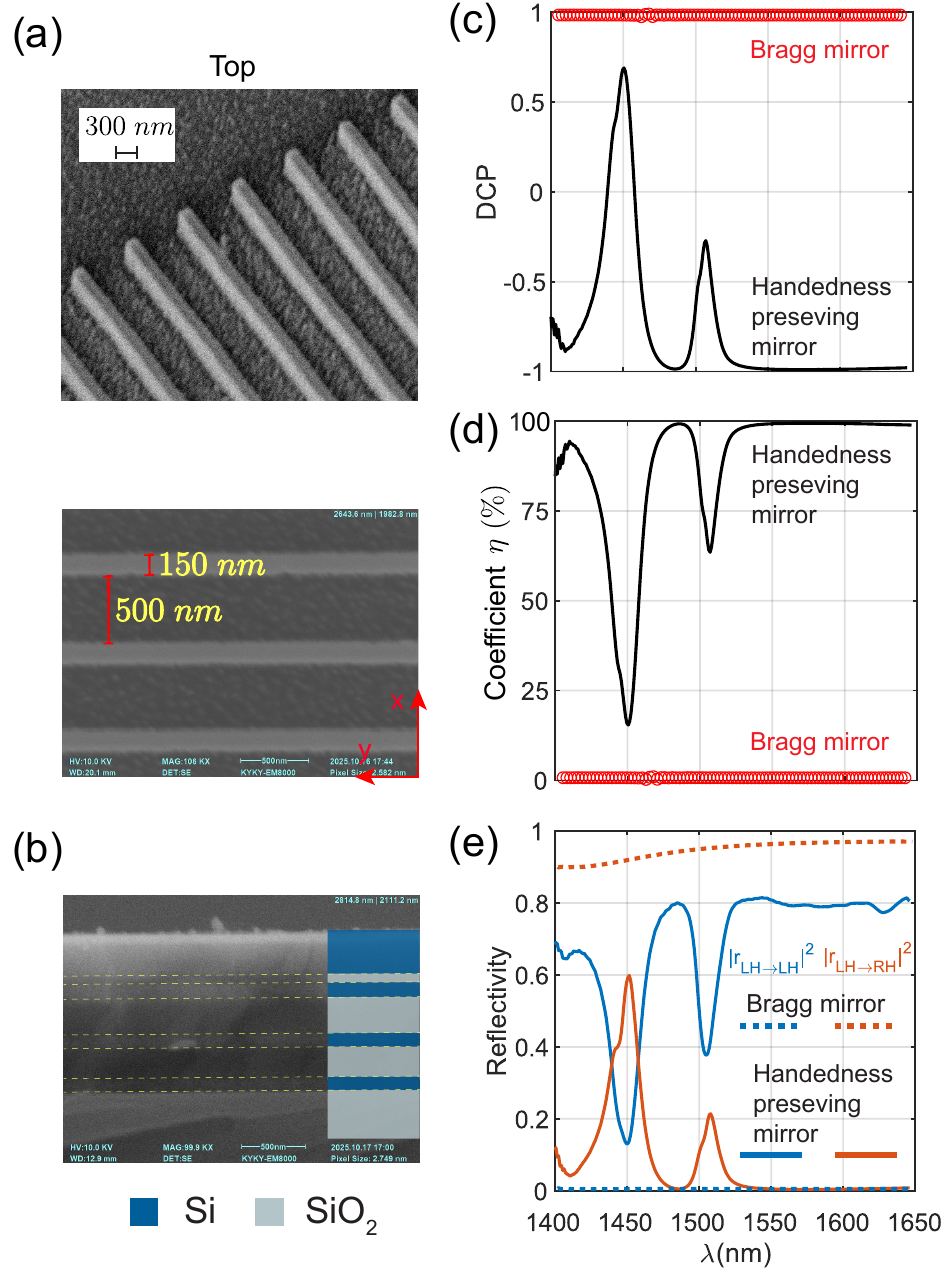}    %UP3.pdf
\caption{Experiment. (a) SEM image of the grating fabricated on top of the Si/SiO$_2$ Bragg mirror. (b) SEM image of the side cross section of the unprocessed Bragg mirror. Measured average parameters of fabricated structure: $h_{gr} =  290$~nm, $h_{\text{SiO}_2}^{\text{sub}} = 60$~nm, $h_{\text{Si}}^{\text{BM}} = 110$~nm, $h_{\text{SiO}_2}^{\text{BM}} = 260$~nm, $p = 650$~nm, $a = 150$~nm, $b = 500$~nm. (c) Spectral dependence of the DCP and (d) handedness-preservation coefficient $\eta$ for the bare Bragg mirror (red markers) and the fabricated handedness-preserving mirror (black line). (e) Co- and cross-handed reflectivity spectra for the Bragg mirror (dashed lines) and the fabricated handedness-preserving mirror (solid lines).}
\label{fig:exp1}
\end{figure}

\begin{figure*}[th!]
\centering\includegraphics[width=1\linewidth]{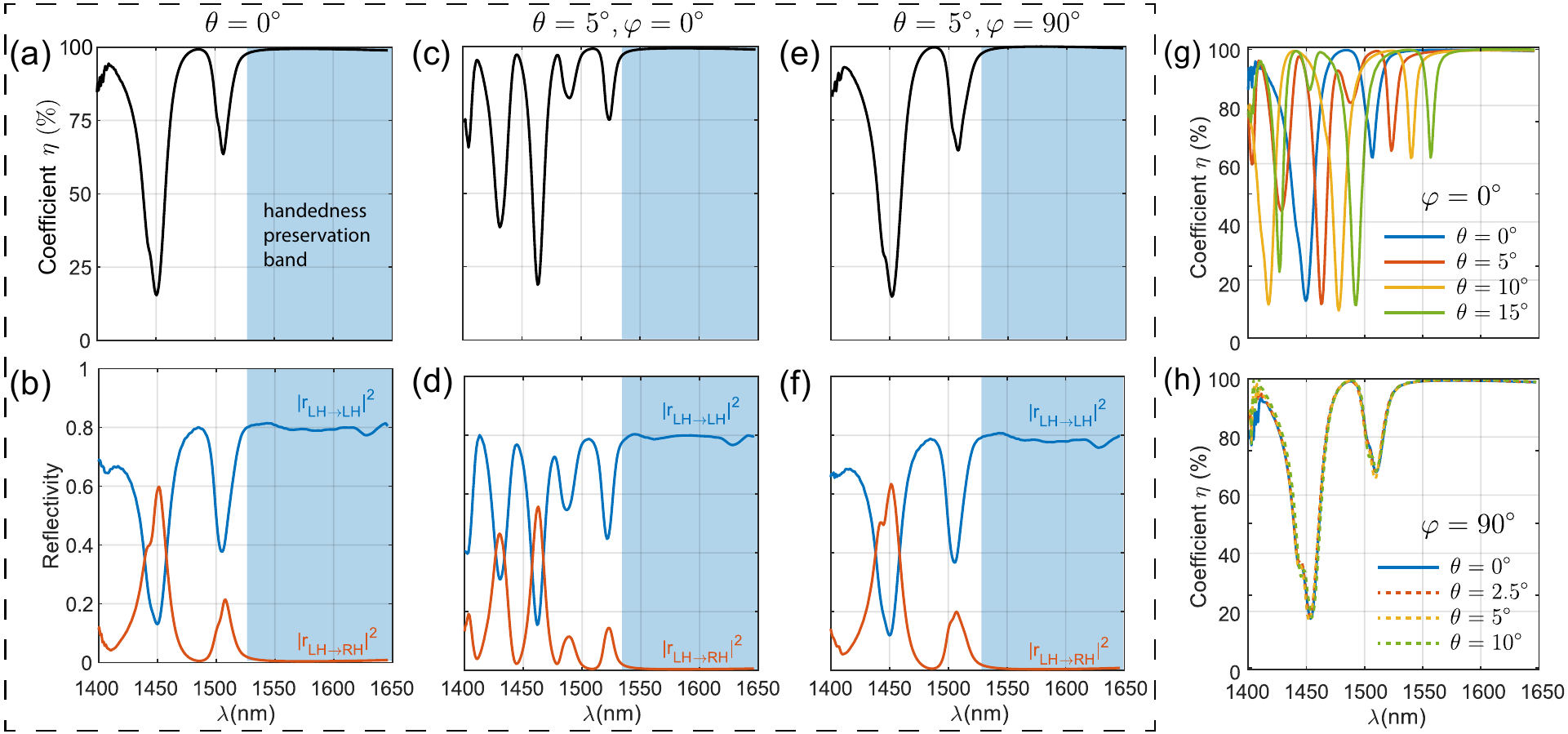}    
\caption{Experiment. Co- and cross-handed reflectivity spectra and the handedness-preservation coefficient $\eta$ for (a,b) normal incidence, (c,d) $\theta = 5^{\circ}$ with the plane of incidence perpendicular to the stripes ($\varphi = 0^{\circ}$), and (e,f) $\theta = 5^{\circ}$ with the plane of incidence parallel to the stripes ($\varphi = 90^{\circ}$).
(g,h) Spectral evolution of the handedness-preservation coefficient $\eta$ for different incidence angles in the two incidence geometries: (g) $\varphi = 0^{\circ}$ and (h) $\varphi = 90^{\circ}$.}
\label{fig:angles}
\end{figure*}

%%%%%%%%%%%%%%%%%%%%%%%%%%%%%%%%%%%%%%%%%%%%%%%%%%%%%%%%%%%%%%%%%%%%%%%%%%%%%%%%%%
%%%%%%%%%%%%%%%%%%%%%%%%%%%%%%%%%%%%%%%%%%%%%%%%%%%%%%%%%%%%%%%%%%%%%%%%%%%%%%%%%%
%%%%%%%%%%%%%%%%%%%%%%%%%%%%%%%%%%%%%%%%%%%%%%%%%%%%%%%%%%%%%%%%%%%%%%%%%%%%%%%%%%
\section{Theory}
\label{sec:theory2}
To describe reflection of a normally incident circularly polarized wave from the mirror, we use the Jones calculus formalism, where the electric field of the wave is expressed in terms of a complex vector of amplitudes written in the linear polarization basis (denoted as "$xy$") or circular polarization basis (denoted as "$\rho\sigma$")
\begin{equation}
    \mathbf{\mathrm{A}}_{xy} = 
    \begin{bmatrix}
        a_x\\a_y
    \end{bmatrix},~~~~~~~
    \mathrm{A}_{\rho\sigma} = 
    \begin{bmatrix}
        a_\rho\\a_\sigma
    \end{bmatrix},
\end{equation}
with a transformation matrix $\mathbb{T}$:
\begin{equation}
    \mathbb{T}\equiv\mathbb{T}_{\rho\sigma \leftarrow xy} = 
    \frac{1}{\sqrt{2}}
    \begin{bmatrix}
        1 & i \\ 1 & -i
    \end{bmatrix},
\end{equation}
such that ${\mathrm{A}}_{\rho\sigma} = \mathbb{T}{\mathrm{A}}_{xy}$. Since the vector of amplitudes can describe waves propagating in either positive or negative $z$-directions, we introduce notation ${\mathrm{A}}^\mathrm{inc}$ for an incident wave, ${\mathrm{A}}^\mathrm{refl}$ for a reflected wave. It should be noted that the vectors ${\mathrm{A}}^\mathrm{inc}_{\rho\sigma} = [1, 0]^T$ and ${\mathrm{A}}^\mathrm{refl}_{\rho\sigma} = [0, 1]^T$ describe a right-handed (RH) wave, while the vectors ${\mathrm{A}}^\mathrm{refl}_{\rho\sigma} = [1, 0]^T$ and ${\mathrm{A}}^\mathrm{inc}_{\rho\sigma} = [0, 1]^T$ correspond to a left-handed (LH) wave. Conventional homogeneous isotropic mirrors reflect both orthogonal polarizations with the same complex amplitude reflection coefficient $r$. The reflection matrices of such a mirror in linear and circular bases are identical:
\begin{equation}
    \mathbb{R}_{xy} = 
    \begin{bmatrix}
        r & 0\\ 0 &r
    \end{bmatrix},~~~
    \mathbb{R}_\mathrm{\rho\sigma} = \mathbb{T}\mathbb{R}_{xy}\mathbb{T}^{-1} = 
    \begin{bmatrix}
        r & 0\\ 0 &r
    \end{bmatrix}.
\end{equation}
Obviously, the reflection matrix $\mathbb{R}_\mathrm{\rho\sigma}$ of isotropic mirrors preserves circular polarization but inverts handedness:
\begin{equation}
    \mathbf{\mathrm{A}}^\mathrm{refl}_{\rho\sigma} \equiv
    \begin{bmatrix}
        r\\0
    \end{bmatrix}^\mathrm{refl} = 
    \begin{bmatrix}
        r & 0\\ 0 &r
    \end{bmatrix}
    \begin{bmatrix}
        1\\0
    \end{bmatrix}^\mathrm{inc} \equiv
    \mathbb{R}_{\rho\sigma}\mathbf{\mathrm{A}}^\mathrm{inc}_{\rho\sigma}.
\end{equation}

\par \setlength\parindent{1em}In contrast to isotropic mirrors, handedness-preserving mirrors invert circular polarization. One can describe this effect by an antidiagonal reflection matrix:
\begin{equation}
    \mathbf{\mathrm{A}}^\mathrm{refl}_{\rho\sigma} \equiv
    \begin{bmatrix}
        0\\r
    \end{bmatrix}^\mathrm{refl} = 
    \begin{bmatrix}
        0 & r'\\ r &0
    \end{bmatrix}
    \begin{bmatrix}
        1\\0
    \end{bmatrix}^\mathrm{inc} \equiv    \mathbb{R}_{\rho\sigma}\mathbf{\mathrm{A}}^\mathrm{inc}_{\rho\sigma}
\end{equation}
Considering only symmetric handedness-preserving mirrors that reflect the LH \textit{and} RH incident waves with \textit{equal} amplitude reflection coefficients ($r = r'$), we obtain reflection matrices in the linear and circular polarization bases in the following form:
\begin{equation}
\label{condr}
    \mathbb{R}_\mathrm{\rho\sigma} = 
    \begin{bmatrix}
        0 & r\\ r &0
    \end{bmatrix},~~~
    \mathbb{R}_{xy} = \mathbb{T}^{-1}\mathbb{R}_\mathrm{\rho\sigma}\mathbb{T} = 
    \begin{bmatrix}
        r & 0\\ 0 &-r
    \end{bmatrix}.
\end{equation}

The expression for the reflection matrix in a linear basis $\mathbb{R}_{xy}$ indicates that preserving the handedness of light upon reflection requires a reflective surface to be designed with two key properties: first, the amplitude reflection coefficients for the $x$ and $y$-polarizations must sum to zero ($r_{xx}+r_{yy} = 0$); second, the cross-polarization coefficients in the linear basis must vanish ($r_{xy} = r_{yx} = 0$). For the surface to act as a high-quality mirror, an additional condition of perfect reflectivity, $|r_{xx}|^2 = |r_{yy}|^2 = 1$, must also be satisfied. In contrast to earlier approaches~\cite{Semnani2020, Voronin2022}, a system meeting these criteria must exhibit in-plane anisotropy while maintaining a vertical plane of mirror symmetry. The simplest structure fulfilling these requirements is a homogeneous slab that is anisotropic within the horizontal plane. While naturally anisotropic materials can, in principle, satisfy these conditions at a specific thickness, their typically weak anisotropy $\Delta n$ results in an impractically narrow operating bandwidth. Periodic metasurfaces, however, can provide substantially stronger in-plane anisotropy. In the following section, we present a metasurface design that simultaneously meets all the above criteria, thereby realizing a broadband, high-quality mirror that preserves the handedness of the reflected light.

\section{Results}
\subsection{Numerical results}
The proposed structure consists of a one-dimensional Si/air grating placed on top of the Si/SiO$_2$ Bragg mirror as shown in Fig.~\ref{fig:mirror}(a). Our goal is to find a set of geometric parameters of the model structure to minimize target function $\mathrm{f}_1 = |r_{xx}+r_{yy}|$ while maximizing target function $\mathrm{f}_2 = |r_{xx}|^2+|r_{yy}|^2$ over the wavelength range between 1400 and 1800~nm. An ideal handedness-preserving mirror corresponds to $\mathrm{f}_1 = 0$ and $\mathrm{f}_2 = 2$. Using genetic algorithm, we optimize the layers' thicknesses $h_{gr}$, $h^\mathrm{sub}_\mathrm{SiO_2}$, the grating period $p$, and the Si wall widths $a$. Thicknesses of the Si and SiO$_2$ layers in the Bragg mirror are fixed and taken such that the corresponding optical thicknesses are equal to the quarter of the wavelength $\lambda=1550$~nm. Introducing a thin SiO$_2$ sublayer between the grating and the top Si layer of the Bragg mirror broadens the operating bandwidth. %The optimal thickness of this sublayer is $h_{\text{SiO}_2}^{\text{sub}} = 50$ nm. Reflection spectra of the structure with optimal geometric parameters are shown in Fig.~\ref{fig:mirror}.
%The optimal parameters are the following (see Fig.~\ref{fig:mirror} (a)): period $p = 503$ nm, width of the stripes $a = 126$ nm, width of the grooves $b = 377$ nm, thickness of the grating $h_{gr} = 297$ nm. The corresponding filling factor $f = a/p \approx 0.25$. For the Bragg mirror (BM), the thicknesses of the Si and SiO$_2$ layers are $h_{\text{Si}}^{\text{BM}} = 102$ nm  and $h_{\text{SiO}_2}^{\text{BM}} = 297$ nm. %\edit{\st{The designed Bragg mirror consists of four Si/SiO$_2$ pairs, whereas the fabricated structure includes only three pairs. Numerical simulations confirm that a three-pair Bragg mirror provides a reflectance of approximately 98\%, in very good agreement with the experimental results}}(see Fig.~\ref{fig:BM}(b)). \edit{\st{ Incorporating four pairs increases the reflectance to 99.5\%. However, this enhancement is not critical for the experiment, as the dominant losses originate from scattering at the grating and fabrication imperfections. }}
%Introducing a thin SiO$_2$ sublayer between the grating and the top Si layer of the Bragg mirror broadens the operating bandwidth. The optimal thickness of this sublayer is $h_{\text{SiO}_2}^{\text{sub}} = 50$ nm.

%\par \setlength\parindent{1em}
The reflection spectra of a structure with the optimized geometric parameters are shown in Fig.~\ref{fig:mirror}. One can see that in the range of 1400--1800~nm, the designed structure simultaneously satisfies requirements $r_{xx}+r_{yy} = 0$ %of a phase difference $\pi$ between the amplitude reflection coefficients $\Delta \phi = \arg(r_{xx})-\arg(r_{yy})$
and $|r_{xx}| = |r_{yy}|=1$ %, that corresponds to $\mathrm{f}_1 = 0$ and $\mathrm{f}_2 = 2$. 
with the cross-polarization amplitude coefficients $r_{xy}$ and $r_{yx}$ equal to zero. As a result, the structure functions as a handedness-preserving mirror within this spectral range, reflecting LH light into LH and RH light into RH, as illustrated in Fig.~\ref{fig:mirror}(d).  

%\par \setlength\parindent{1em}
The operating bandwidth of the mirror is governed primarily by the spectral range over which the condition $r_{xx}+r_{yy} = 0$ is satisfied, rather than by the reflectance band of the underlying Bragg mirror; the latter is substantially broader. In Fig.~\ref{fig:mirror}(d), in addition to a wide handedness-preservation band, several sharp dips are observed near 1200~nm. These parasitic resonant features originate from the excitation of waveguide modes within the silicon layers of the Bragg mirror, mediated by diffraction from the grating. Under oblique incidence, these resonances split due to the lifting of polarization degeneracy.

As can be seen from Fig.~\ref{fig:mirror}, the effect of handedness preservation is non-resonant since it is wideband and falls entirely within the Bragg-mirror's stop-band. This indicates that the underlying mechanism does not rely on sharp, narrowband resonances but rather on the engineered anisotropic properties of the effective medium.

%%%%%%%%%%%%%%%%%%%%%%%%%%%%%%%%%%%%%%%%%%%%%%%%%%%%%%%%%%%%%%%%%%%%%%%%%%%%%%%%%%%%%%%%%%%%%%%%%%%%%%%%%%%%%%%%%%

Due to the non-resonant nature of the handedness-preserving functionality, the designed mirror should be stable against geometric imperfections. Below, we analyze the tolerance of the structure to geometric fabrication imperfections across all geometrical parameters. We begin with the assessment of the grating period, as shown in Fig.~\ref{fig:geom}(a). Increasing the period slightly narrows the operating bandwidth and shifts it toward longer wavelengths. A variation of up to 50~nm preserves the reflectance above 90\% across the entire band from 1400~nm to 1800~nm. The most sensitive parameter is the filling factor (see Fig.~\ref{fig:geom}(b)), as it directly determines the in-plane anisotropy of the layer and, consequently, the phase retardance between the $r_{xx}$ and $r_{yy}$ components. Nevertheless, the reflection is maintained above 90\% for variations in the filling factor within $\pm15\%$ that correspond to the change of $\pm75$~nm in the widths of the grooves and the stripes. The dependence on the grating thickness $h_{\text{gr}}$ (Fig.~\ref{fig:geom}(c)) is qualitatively similar to that on the period (Fig.~\ref{fig:geom}(c)). The response remains highly robust, with the reflectance remaining above 90\% throughout the band. The operating range shifts slightly toward longer wavelengths due to a minor increase in the phase difference between the $r_{xx}$ and $r_{yy}$ components $\Delta\phi$ as the grating thickness increases. As shown in Fig.~\ref{fig:geom}(d), the broadest spectral range with reflectance exceeding 99\% is achieved at the optimal SiO$_2$ sublayer thickness. Nevertheless, for all sublayer thicknesses presented in Fig.~\ref{fig:geom}(d), the reflectance remains above 90\% across the entire spectral band. Notably, even in the absence of the SiO$_2$ sublayer, the structure continues to function as a handedness-preserving mirror, albeit with approximately half the operating bandwidth. Figure~\ref{fig:geom}(e) shows the reflection coefficient under simultaneous variations of the Si and SiO$_2$ layer thicknesses in the Bragg mirror. A thickness deviation of $\pm10$~nm does not reduce the width of the operating band, and even a deviation of $\pm20$~nm maintains reflection above 90\% across the entire band.

%\par \setlength\parindent{1em}
Another important characteristic of the designed mirror is its stability with respect to the angle of incidence. 
%\edit{\st{For a one-dimensional grating, such as the one considered in this work, there are two fundamentally different configurations: the incidence plane is perpendicular to the grating stripes $\varphi = 0^{\circ}$, and the incidence plane is parallel to the stripes $\varphi = 90^{\circ}$.}} Figure~\ref{fig:angle}\edit{\st{reveals that the two excitation geometries exhibit qualitatively different behaviors of the sharp resonances on the short-wavelength side of the operating band that in detail discussed in Supplementary materials.}}
Figure~\ref{fig:angle} shows the angular dependence of the co-polarized reflection spectrum in the circular basis. When the plane of incidence is perpendicular to the stripes (left panel of Fig.~\ref{fig:angle}), the waveguide resonances redshift with increasing incident angle, which restricts the operating bandwidth of the mirror. However, outside these resonances, the reflectance remains high in a wide spectral range, reaching up to 99\% for incidence angles below $20^{\circ}$ and remaining above 90\% for angles up to $50^{\circ}$. In contrast, when the plane of incidence is parallel to the stripes (right panel of Fig.~\ref{fig:angle}), no resonant features appear in the spectral–angular region of interest, which ensures a high stability of the mirror performance with respect to the incident angle. As a result, the reflectance remains high over the entire 1400--1800~nm range, exceeding 99\% for angles below $20^{\circ}$ and staying above 90\% for angles up to $50^{\circ}$.

%%%%%%%%%%%%%%%%%%%%%%%%%%%%%%%%%%%%%%%%%%%%%%%%%%%%%%%%%%%%%%%%%%%%%%%%%%%%%%%%%%%%%%%%%%%%%%%%%%%%%%%%%%%%%%%%%%
\subsection{Experimental results}
To verify our theoretical findings, we fabricated a sample of the designed metasurface %The grating was patterned by etching the silicon layer atop a Bragg mirror; the Bragg mirror itself consists of three Si/SiO$_2$ pairs deposited on a quartz substrate. 
on the Bragg mirror consisting of three Si/SiO$_2$ pairs. Details of the sample fabrication procedure are provided in the Materials and Methods section. Figure~\ref{fig:exp1}(a,b) shows a scanning electron microscope (SEM) image of the fabricated grating and Bragg mirror.

To quantify the preservation of optical handedness upon reflection from the fabricated mirror, we measured the reflected intensities for both circular polarizations and calculated the degree of circular polarization (DCP):
\begin{equation*}
    \text{DCP} = \frac{I(\mathbf{\sigma^-}) - I(\mathbf{\sigma^+})}{I(\mathbf{\sigma^-}) + I(\mathbf{\sigma^+})},   
\end{equation*}
where $\mathbf{\sigma^-}$ and $\mathbf{\sigma^+}$ denote circular polarization states corresponding to the Jones vectors $\mathrm{A}_{\rho\sigma} = [0,1]^T$ and $\mathrm{A}_{\rho\sigma} = [1,0]^T$ respectively (see Sec.~Theory). The experimental setup and measurement procedure are detailed in the Materials and Methods section and schematically illustrated in Fig.~S4.

In addition to the DCP, we introduce a handedness preservation coefficient $\eta$, which is better tailored for the quantitative description of handedness preservation upon reflection. This coefficient can be expressed in terms of DCP or through the amplitudes of co- and cross-handed reflection coefficients as:
\begin{equation*}
    \eta = (1-\text{DCP})/2\cdot100\%= \frac{|\text{r}_{_{\text{LH}\rightarrow\text{LH}}}|^2}{|\text{r}_{_{\text{LH}\rightarrow\text{LH}}}|^2+|\text{r}_{_{\text{LH}\rightarrow\text{RH}}}|^2}\cdot100\%.
\end{equation*} 

In accordance with Sec.~Theory, the incident $\sigma^{-}$ and reflected $\sigma^{+}$ polarization states correspond to LH light, whereas the incident $\sigma^{+}$ and reflected $\sigma^{-}$ states correspond to RH light. Given these conventions, illumination with $\sigma^{-}$ light yields $\mathrm{DCP} = +1$ and $\eta = 0$ when the polarization state is conserved upon reflection, while the handedness is not conserved. Conversely, $\mathrm{DCP} = -1$ and $\eta = 1$ indicate conversion from $\sigma^{-}$ to $\sigma^{+}$ polarization while preserving the handedness.

For a Bragg mirror without a grating, the reflected light remains in the $\mathbf{\sigma^-}$ polarization state following the DCP = 1 and the handedness preservation coefficient $\eta = 0$ at all wavelengths within the range 1400--1650~nm, as shown by the red markers in Fig.~\ref{fig:exp1}(c)--(d). At the same time, when light is reflected from the Bragg mirror capped by the Si stripes, the polarization $\mathbf{\sigma^-}$ converts to $\mathbf{\sigma^+}$, and the handedness of the light remains the same. In that case, in the spectral range over 1525~nm, the DCP is close to -1, with the preservation coefficient $\eta $ reaching 99\%. In the following discussion, we do not consider wavelengths below 1400~nm because this region is beyond the operating band of the mirror.

Fig.~\ref{fig:exp1}(e) presents the cross- and co-handedness reflectivity of circularly polarized light, defined as the ratio of the reflected intensity with a given handedness to the total incident intensity. For the unprocessed Bragg mirror, the reflection of LH light occurs primarily in the RH channel, as shown by the orange dashed line, while the reflection of LH light in the LH channel (blue dashed line) is approximately zero across the considered spectral range; the total reflectance is approximately 98\%. In contrast, the fabricated handedness-preserving mirror reflects LH light primarily in the LH channel, as demonstrated by the solid lines, achieving this functionality around the wavelength of 1525~nm.

To test the stability of the handedness-preservation effect with respect to the incident angle $\theta$, we measured the reflection spectra for angles up to $15^\circ$ in planes parallel and perpendicular to the stripes (the $xz$ and $yz$ planes, respectively). Under oblique incidence within the $xz$ plane ($\varphi = 0^{\circ}$), the parasitic waveguide-related resonances split and exhibit an additional red shift. At an incidence angle of $\theta = 5^{\circ}$, the handedness-preservation band spans from 1538~nm (Fig.~\ref{fig:angles}(c)), while the LH-to-LH reflectivity within this band remains between 78\% and 80\% (Fig.~\ref{fig:angles}(d)). In contrast, for incidence within the $yz$ plane, the resonances neither split nor undergo a noticeable spectral shift. At $\theta = 5^{\circ}$, the handedness-preservation band begins at 1535~nm (Fig.~\ref{fig:angles}(e)), with the co-handed reflectivity remaining between 80\% and 81\% (Fig.~\ref{fig:angles}(f)).

The extended angular dependence is presented in Fig.~\ref{fig:angles}(g)-(h). For excitation in the plane perpendicular to the stripes ($\varphi = 0^{\circ}$), the spectra contain a handedness-preserving band for all incidence angles within the range $|\theta| \leq 15^{\circ}$; however, this band shifts to longer wavelengths, as shown in Fig.~\ref{fig:angles}(g). As an example, the green line in Fig.~\ref{fig:angles}(g) shows the spectral dependence of the handedness-preservation coefficient for $\theta = 15^{\circ}$ and $\varphi = 0^{\circ}$; a well-defined band is still observed but is redshifted to a range extending up to 1575~nm. For excitation in the plane parallel to the stripes ($\varphi = 90^{\circ}$), the spectra exhibit almost no sensitivity to the incidence angle, as seen in Fig.~\ref{fig:angles}(h).

These observations demonstrate that the handedness-preserving functionality of the mirror is robust against moderate angular deviations and remains effective across both excitation geometries, confirming the stability of the design under different illumination conditions.
%%%%%%%%%%%%%%%%%%%%%%%%%%%%%%%%%%%%%%%%%%%%%%%%%%%%%%%%%%%%%%%%%%%%%%%%%%%%%%%%%%%%%%%%%%%%%%%%%%%%%%%%%%%%%%%%%%

\section*{Discussions}

Next, let us describe the above effect by considering a model structure consisting of an anisotropic layer with an effective dielectric permittivity tensor $\hat\varepsilon = diag [\varepsilon_o, \varepsilon_e, \varepsilon_o]$, lying on a perfect mirror ($\varepsilon_m=-\infty$). For this model structure, the amplitude reflection coefficients $r_{xx}$ and $r_{yy}$ for a normally incident plane wave can be expressed as:
\begin{align}
    r_{xx} &= \frac{1-i\sqrt{\varepsilon_e}\cot{\left(k_0\sqrt{\varepsilon_e}h_{gr}\right)}}{1+i\sqrt{\varepsilon_e}\cot{\left(k_0\sqrt{\varepsilon_e}h_{gr}\right)}},\\
    r_{yy} &= -\frac{1+\frac{i}{\sqrt{\varepsilon_o}}\tan{\left(k_0\sqrt{\varepsilon_o}h_{gr}\right)}}{1-\frac{i}{\sqrt{\varepsilon_e}}\tan{\left(k_0\sqrt{\varepsilon_o}h_{gr}\right)}}
\end{align}
The condition for the anisotropic layer to function as a handedness-preserving mirror, $r_{xx}+r_{yy} = 0$, is now equivalent to 
\begin{equation}
\label{effc}
    \frac{\tan{\left(k_0\sqrt{\varepsilon_e}h_{gr}\right)}}{\sqrt{\varepsilon_e}}\cdot\frac{\tan{\left(k_0\sqrt{\varepsilon_o}h_{gr}\right)}}{\sqrt{\varepsilon_o}} = -1.
\end{equation}
By taking effective dielectric permittivities from the FMM-based homogenization procedure as $\varepsilon_o = 1.40$ and $\varepsilon_e = 5.37$, one can easily check that condition~\eqref{effc} is approximately satisfied within the handedness-preservation band.

As mentioned in Sec.~Theory, one of the alternatives to implement a reflective surface such that its reflection matrix takes the form \eqref{condr} is the use of a homogeneous anisotropic layer with in-plane anisotropy. Although achieving a constant phase shift of $\pi$ between the reflection coefficients $r_{xx}$ and $r_{yy}$ is possible at a single wavelength, maintaining this condition over a broad spectrum while also ensuring that both reflection amplitudes remain near unity is exceptionally demanding. It necessitates an anisotropic material with anomalous dispersion of its ordinary and extraordinary permittivities and, moreover, with a constant difference between them (see Supplemental Material, Fig.~S5). Due to such a specific requirement, the successful selection of a suitable anisotropic material appears highly unlikely.polarization, which are commonly employed for detection of chiral molecules.

Multilayer stacks of anisotropic materials may offer  another alternative solution. Following the design proposed by Pancharatnam \cite{pancharatnam1955achromatic}, commercial quarter- and half-wave plates with extremely flat retardance in transmission over a very broad spectral range have been demonstrated. An example is “Thorlabs Superachromatic Wave Plates,” which consist of three quartz and three magnesium fluoride (MgF$_2$) plates that are optically cemented to maximize transmission and are carefully aligned to minimize the wavelength dependence of retardance \cite{thorlabs11}. A combination of such a broadband quarter-wave plate with an integrated mirror deposited on one of its surfaces can function as a handedness-preserving mirror. To compare its characteristics with those of our mirror, we simulated the above-mentioned quarter-wave plate placed in front of a broadband metallic mirror (for details, see the Supplementary Material and Fig.~S6(a)--(b)). As expected, this approach provides extremely broadband operation, with a spectral bandwidth approximately five times larger than that of our design. However, it does not offer advantages in reflectivity, showing a reflectance of only about 80\%, which is similar to the experimentally achieved values and significantly lower than the theoretically predicted values for our design. Unlike our structure, the non-perfect reflectance of the quarter-wave plate-based mirror is caused by, among other things, losses (according to \cite{thorlabs11} — the sum of reflection and transmission coefficients is less than 1) that sufficiently suppress the quality factor of such a mirror. Moreover, in comparing both designs of a handedness-preserving mirror, it is worth mentioning that our grating-based design exhibits a substantial advantage in stability with respect to the angle of incidence. The quarter-wave plate based mirror works primarily for normal incidence and shows a rapid degradation of retardance for incidence angles exceeding $3^{\circ}$. As mentioned above, numerical simulations show that in our design, the mirror is nearly insensitive to variations in the angle of incidence within $\pm 20^{\circ}$. A detailed comparison between handedness-preserving mirrors based on commercial wave plates and our design is presented in the Supplemental Material (see Fig.~S6(c)-(f)). An additional advantage of our design over mirrors based on commercial quarter-wave plates is its compactness. The thickness of the grating is 297~nm, and the total thickness of the structure that includes the Bragg mirror is approximately 2~$\mu$m. In contrast, the thickness of a mirror based on a quarter-wave plate exceeds 5~mm. Moreover, our structure is fully compatible with integrated circuit fabrication technologies and can be manufactured from a wide variety of materials.

An alternative design of a Bragg mirror-based handedness-preserving mirror can be realized by replacing the dielectric Bragg mirror with a metallic mirror, as the latter can provide a 97\% reflectance and a high angular tolerance. In the Supplemental materials, we demonstrated that an optimized geometry of the metallic mirror-based design supports handedness-preserving reflection over a spectral range from 1350~nm to 2100~nm (Fig.~S3). Although this operating bandwidth is approximately 300~nm wider than that of the Bragg-mirror-based design, the co-handed reflectivity varies between 95\% and 97\%, as shown in Fig.~S3(d), which is less than the 99\% offered by the configuration considered in Sec.~Results. 

As mentioned in Sec.~Introduction, one of the applications of handedness preserving mirrors is their use in the creation of Fabry-Perot resonators that support chiral electromagnetic modes \cite{dyakov2024chiral,dyakov2025strong}. For this purpose, one or both mirrors should be transparent. In this perspective, handedness-preserving mirrors based on commercial quarter-wave plates \cite{thorlabs333} as well as those on reflective metallic substrates are inappropriate candidates since they absorb light.

A Fabry-Perot resonator formed by anisotropic broadband handedness-preserving mirrors is shown in Fig.~S7. As shown in our recent publications \cite{dyakov2024chiral, dyakov2025strong}, such a resonator supports modes of both handedness; by choosing the appropriate gap size and the angle between mirror orientations, one can easily adjust the frequency of the Fabry-Perot mode to any value within the handedness-preservation band of the mirrors. This capability fundamentally distinguishes Fabry-Perot resonators based on anisotropic handedness-preserving mirrors from those utilizing low-symmetric handedness-preserving mirrors \cite{Semnani2020, Voronin2022}, as the latter operate only within a narrow spectral range. As mentioned, the narrow band functionality of low-symmetric mirrors originates from their resonant nature. 

Finally, as demonstrated in ~\cite{dyakov2025strong}, the resonant frequency of the Fabry-Perot resonator based on a pair of anisotropic handedness-preserving mirrors is highly sensitive to the macroscopic dielectric parameters of the material placed between the mirrors. The presence of chiral modes over a wide spectral range allows for chiral sensing by monitoring the resonance frequency, without the need to measure circular dichroism or optical activity.

\section*{Conclusion}
In conclusion, we have theoretically proposed and experimentally realized a wideband, all-dielectric mirror capable of preserving the handedness of circularly polarized light upon reflection in the near-infrared range. By combining a high-contrast, near-subwavelength one-dimensional dielectric grating with a Bragg mirror and optimizing the design via a genetic algorithm, we achieved a structure that is robust against fabrication imperfections and tolerant to oblique incidence. Our experimental results demonstrate a wide spectral band in which over 98\% of the reflected light preserves its handedness, with a total reflection coefficient reaching 80\% under normal incidence. Moreover, the mirror retains high performance for incidence angles up to ±15$^\circ$, underscoring its practical robustness. These findings demonstrate that the proposed mirror can effectively function as a reflective phase plate, providing a promising platform for Fabry–Pérot resonators and other photonic devices engineered for chiral light control and manipulation.

\section*{Materials and Methods}
\subsection{Sample Fabrication}

The handedness-preserving mirrors are fabricated on a single-crystal quartz substrate. A Bragg mirror consisting of three consecutive pairs of Si and SiO$_2$ layers is first deposited on the bare substrate using thermal e-gun deposition, with layer thicknesses of Si (110~nm) / SiO$_2$ (280~nm) / Si (110~nm) / SiO$_2$ (280~nm) / Si (110~nm). A 60~nm thick SiO$_2$ layer and a 290 nm thick Si layer are subsequently deposited on top to form the grating layer. Next, an aluminum (Al) mask defining the grating pattern is fabricated using UV lithography and thermal e-gun deposition of a 100~nm thick Al film. This Al mask serves as an etch mask for the subsequent inductively coupled plasma reactive ion etching (ICP RIE) of the top Si layer in an SF$_6$/CHF$_3$ gas mixture. Finally, the Al mask is dissolved in a weak base solution, leaving a clean Si grating. The final structure covers an area of $4\times4$ mm$^2$, with a grating period of approximately 650~nm and a ribbon thickness of approximately 150~nm (see Fig.~\ref{fig:exp1}(a)).

\subsection*{Measurement procedure}
The experimental results reported in this work were obtained using a resonant polarization-reflection technique. The experimental scheme is shown in Fig.~S4. All measurements were performed at room temperature. 

The broadband light of a halogen lamp was directed onto a 500-µm pinhole, and the resulting image was projected onto the sample by a lens-based optical system. Before reaching the sample, the beam passed through an interference filter (IF, LBTEK) that selected wavelengths in the 1100–2100 nm range. A Glan–Taylor polarizer (GP1, Thorlabs) combined with a quarter-wave Fresnel rhomb retarder (qwFR, Thorlabs) was used to generate the required circular polarization state of the incident beam.

An iris diaphragm positioned in front of a plate beamsplitter (PB, Thorlabs, 10:90) limited the beam aperture. The beamsplitter, mounted on a micrometer translation stage, enabled lateral displacement of the beam along the X-axis and thus controlled the angle of incidence. The sample, a handedness-preserving mirror, was placed on a precision XYZ translation stage with the possibility of rotation around the Z axis, and the position of the focal spot on its surface was monitored using an optical camera (The Imaging Source).

The reflected signal passed back through a Fresnel rhomb retarder and a Glan–Taylor polarizer (GP2, Thorlabs), allowing for the measurement of the intensity distribution in the circular polarization basis. The beam then passed through an achromatic quarter-wave plate (qwP) and was subsequently focused onto the entrance slit of the spectrometer. A quarter-wave plate was used to suppress any residual polarization effects introduced by the monochromator grating in the spectrometer. Spectral measurements were performed using an Andor SR-303i-B spectrometer coupled to a thermoelectrically cooled CCD camera (Andor DU 490A-1.7). The exposure time was 10~s, with count rates below 300~Kcps. The spectral resolution was 1.5~nm, and the error in measuring the DCP signal did not exceed 0.5\%.

\section*{Acknowledgement}
This work was supported by the Russian Science Foundation (Grant No. 25-12-00454).

\section*{Authors contributions}
 S.D. conceived and designed the research. N.S. carried out the numerical simulations and theoretical analysis. O.K. developed the fabrication process and manufactured the samples. A.D. and V.K. designed and performed the experimental measurements. N.S. and A.D. analyzed and interpreted the data.
The manuscript was mainly written by N.S., S.D. and A.D., who made the most significant contribution to drafting, structuring, and revising the paper. All authors discussed the results and approved the final version of the manuscript.

\section*{Conflict of interest}

The authors declare no conflicts of interest.

%%%%%%%%%%%%%%%%%%%%%%%%%%%%%%%%%%%%%%%%%%%%%%%%%%%%%%%%%%%%%%%%%%%%%%%%%%%%%%%%%%%%%%%%%%%%%%%%%%%%%%%%%%%%%%%%%%
\bibliography{biblio}

%%%%%%%%%%%%%%%%%%%%%%%%%%%%%%%%%%%%%%%%%%%%%%%%%%%%%%%%%%%%%%%%%%%%%%%%%%%%%%%%%%%%%%%%%%%%%%%%%%%%%%%%%%%%%%%%%%
%%%%%%%%%%%%%%%%%%%%%%%%%%%%%%%%%%%%%%%%%%%%%%%%%%%%%%%%%%%%%%%%%%%%%%%%%%%%%%%%%%%%%%%%%%%%%%%%%%%%%%%%%%%%%%%%%%
%%%%%%%%%%%%%%%%%%%%%%%%%%%%%%%%%%%%%%%%%%%%%%%%%%%%%%%%%%%%%%%%%%%%%%%%%%%%%%%%%%%%%%%%%%%%%%%%%%%%%%%%%%%%%%%%%%
%%%%%%%%%%%%%%%%%%%%%%%%%%%%%%%%%%%%%%%%%%%%%%%%%%%%%%%%%%%%%%%%%%%%%%%%%%%%%%%%%%%%%%%%%%%%%%%%%%%%%%%%%%%%%%%%%%
%%%%%%%%%%%%%%%%%%%%%%%%%%%%%%%%%%%%%%%%%%%%%%%%%%%%%%%%%%%%%%%%%%%%%%%%%%%%%%%%%%%%%%%%%%%%%%%%%%%%%%%%%%%%%%%%%%

\clearpage
\setcounter{figure}{0}           
\renewcommand{\thefigure}{S\arabic{figure}}

\section*{Supplementary information: \\ Broadband wide-view all dielectric handedness preserving mirror}

\begin{figure*}[t!]
\centering\includegraphics[width=0.7\linewidth]{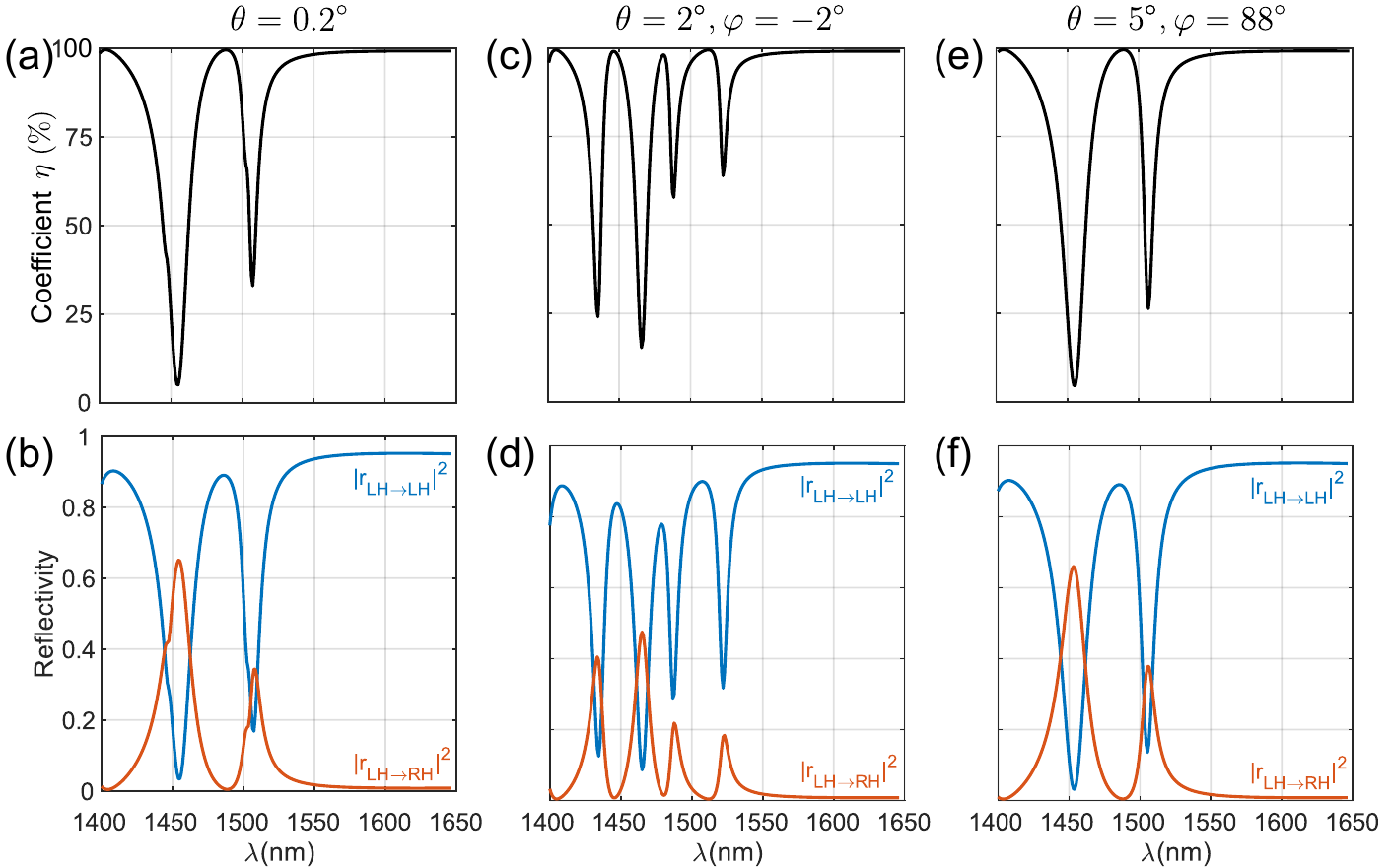}   
\caption{Modeling. Spectra of (a),(c),(e) handedness preservation coefficient and (b),(d),(f) reflectivity of handedness preserving mirror. Reproduction of the experimental result from Fig.~5 (a-f). Parameters of the structure used for modeling: $p = 644$ nm, $a = 154.5$ nm, $b = 489.5$ nm, $h_{gr} = 290$ nm, $h_{\text{SiO}_2}^{\text{sub}} = 110$ nm, two upper layers of Si in Bragg mirror have thicknesses $h_{\text{Si}}^{\text{DBR}} = 97$ nm, and the lower - $h_{\text{Si}}^{\text{DBR}} = 102$ nm,  $h_{\text{SiO}_2}^{\text{DBR}} = 260$ nm.}
\label{fig:mod}
\end{figure*}

\begin{figure*}[th]
\centering\includegraphics[width=1\linewidth]{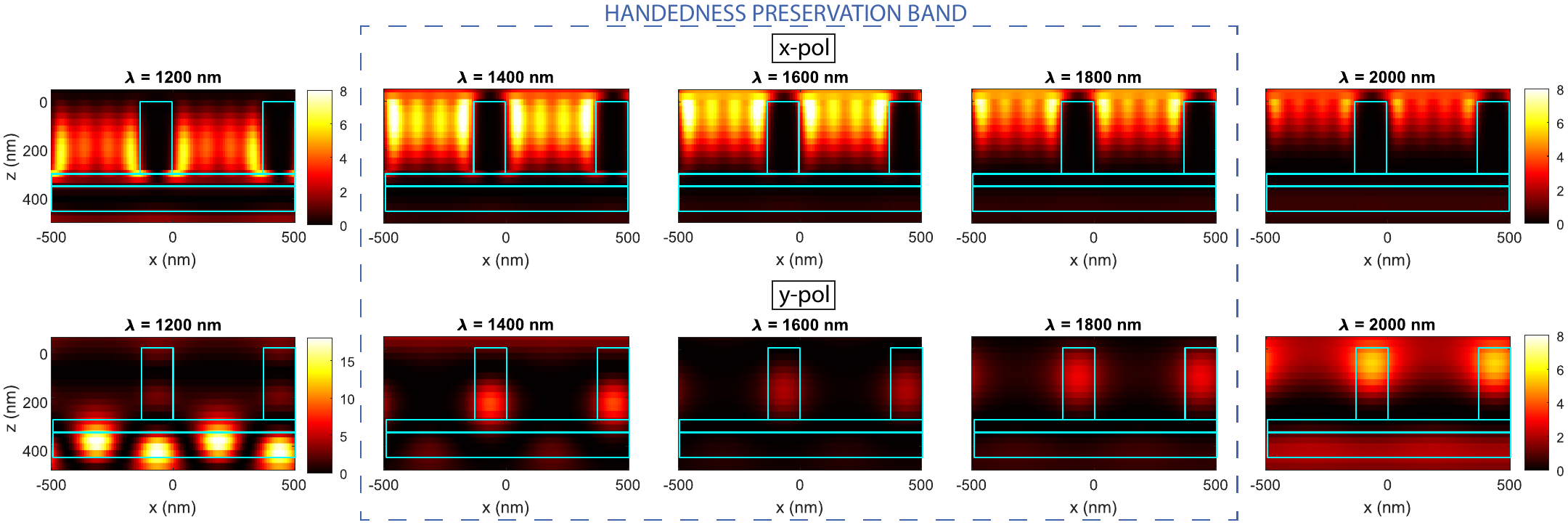}    
\caption{Electric field intensity $|E|^2$ distributions in the $x$–$z$ cross section for (top row) $x$-polarized and (bottom row) $y$-polarized incident light. Different rows correspond to different wavelengths, illustrating the difference in field distributions inside the operating band and at shorter and longer wavelengths.}
\label{fig:fields}
\end{figure*}

\begin{figure*}[t!]
\centering\includegraphics[width=1\linewidth]{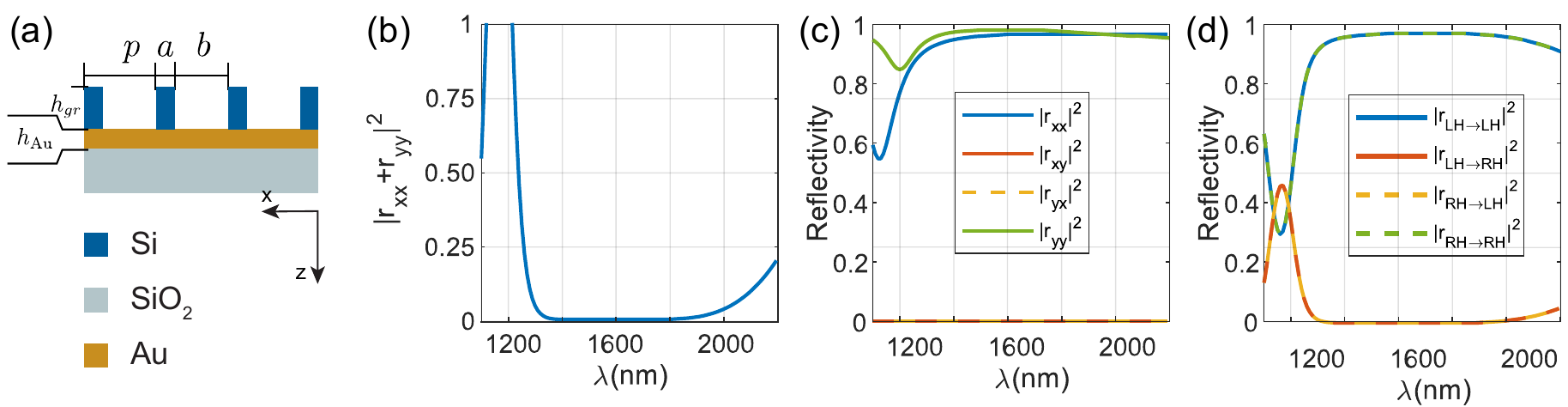}    
\caption{Modeling. (a) The sketch of a wide-band handedness-preserving mirror consisting of the one-dimensional Si grating on a gold mirror. (b) Spectral dependence of $|r_{xx}-r_{yy}|^2$ which characterize the difference of amplitudes and phases. (c) Spectral dependencies of cross-polarization reflection coefficients in the basis of linear polarizations. (d) Spectral dependencies of cross-polarization reflection coefficients in the basis of circular polarizations. Parameters used for modeling: thicknesses of the grating is $h_{gr} = 332$~nm,  the period $p = 646$~nm, the stripes' width $a = 142$~nm, the grooves' width $b = 504$~nm and the thickness of gold layer $h_{\text{Au}} = 183$~nm.}
\label{fig:metalHPM}
\end{figure*}

\begin{figure*}[h!]
\centering\includegraphics[width=0.7\linewidth]{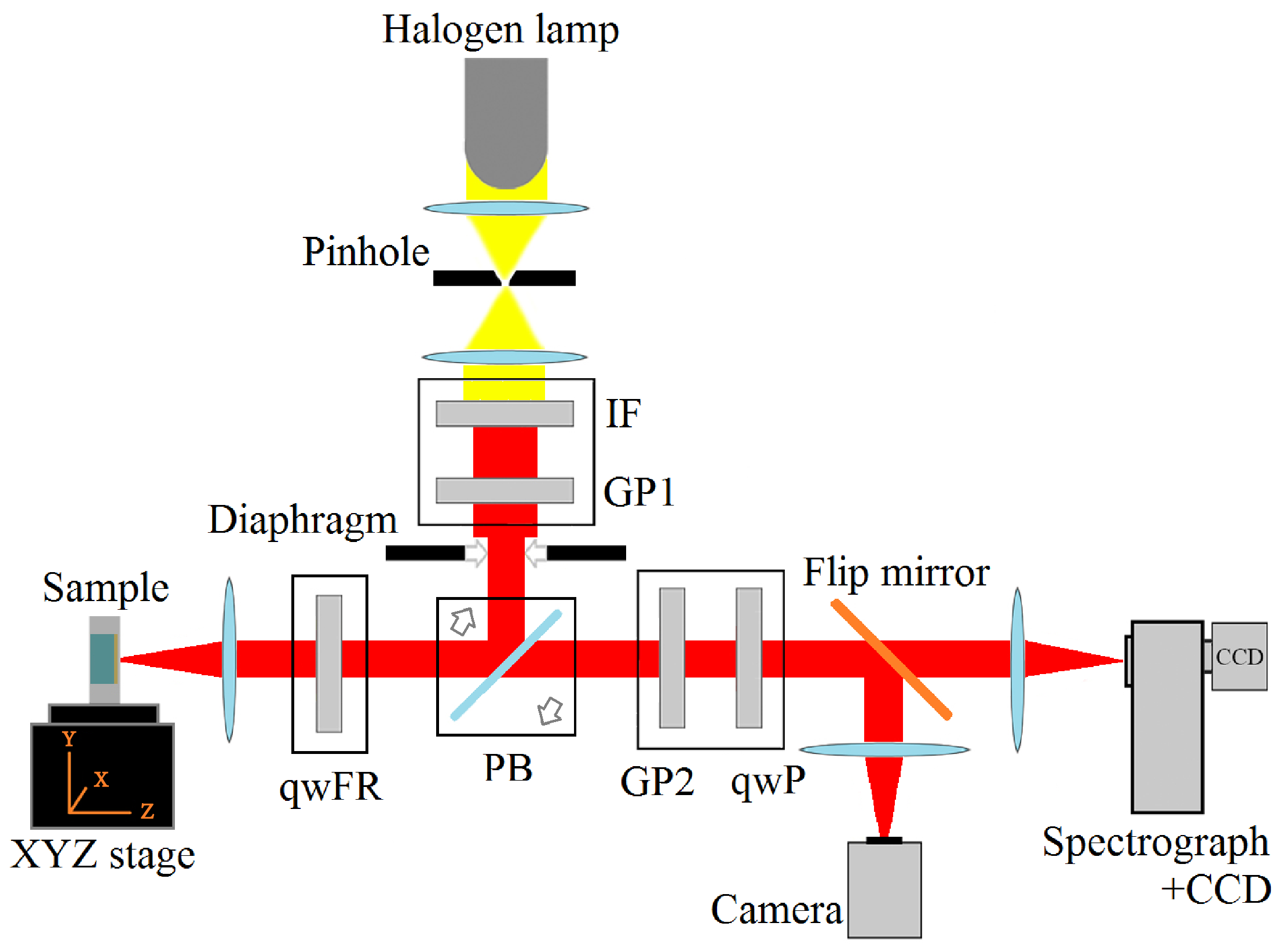}  
\caption{The scheme of optical measurements of resonant reflection from the sample. Broadband unpolarized light from a Halogen lamp passed through an interference filter (IF) and was converted into circularly polarized light using a combination of a Glan–Taylor polarizer (GP1) and a quarter-wave Fresnel rhomb retarder (qwFR). A quarter-wave Fresnel rhomb retarder was used because it provides strongly uniform retardation $\lambda/4$ in a broad spectral range. The beam was then focused onto the sample. An iris diaphragm (Diaphragm) was used to control the incident beam aperture before a plate beamsplitter (PB). A beamsplitter was mounted on a micrometer translation stage, allowed lateral displacement of the beam along the X-axis, enabling measurements of the angular dependence of the resonant reflection. The circular polarization of the signal reflected from the handedness-preserving mirror was analyzed using the same Fresnel rhomb retarder (qwFR) in combination with a Glan–Taylor polarizer (GP2). An achromatic quarter-wave plate (qwP) was inserted to suppress any polarization effects introduced by the monochromator grating. The reflected signal was then focused onto the entrance slit of a spectrometer and detected using a CCD camera. The sample position was controlled by visual monitoring of the focal spot with an optical camera (Camera).}
\label{fig:setup}
\end{figure*}
%%%%%%%%%%%%%%%%%%%%%%%%%%%%%%%%%%%%%%%%%%%%%%%%%%%%%%%%%%%%%%%%%%

\subsection*{Incidence angle dependence}

Figure~3 in the main text reveals that the two excitation geometries exhibit qualitatively different behaviors of the sharp resonances on the short-wavelength side of the operating band. The excitation of these resonances requires the fulfillment of the in-plane momentum conservation condition, $k_{\parallel} = \beta$, where $\beta$ is the propagation constant of the guided mode supported by the Si layers of the Bragg mirror. The diffracted waves generated by the grating possess in-plane wave vectors
\begin{equation*}
\mathbf{k}_{\parallel,n} =
\begin{pmatrix}
k_{x,n} \\
k_{y,n}
\end{pmatrix},
\end{equation*}
with
\begin{align*}
k_{y,n} &= k_y = \frac{2\pi}{\lambda}\sin\theta\sin\varphi, \\
k_{x,n} &= \frac{2\pi}{\lambda}\sin\theta\cos\varphi + n\frac{2\pi}{p}, \quad n\in\mathbb{Z}.
\end{align*}

In the configuration with $\varphi = 90^{\circ}$, the magnitude of the in-plane wave vector is given by
\begin{equation}
\label{ky}
k_{\parallel,n} = \sqrt{\left(\frac{2\pi}{\lambda}\sin\theta\right)^2 + \left(n\frac{2\pi}{p}\right)^2}.
\end{equation}
Thus, the requirement of in-plane momentum conservation for any fixed diffraction order $n$ implies that an increase in the incidence angle $\theta$ leads to a shift of the resonance toward shorter wavelengths, as observed in Fig.~3 (right).

For the configuration with $\varphi = 0^{\circ}$, the in-plane wave-vector magnitude becomes
\begin{equation}
\label{kx}
k_{\parallel,n} = \sqrt{\left(\frac{2\pi}{\lambda}\sin\theta\right)^2 + 2 n \frac{2\pi}{\lambda}\sin\theta \frac{2\pi}{p} + \left(n\frac{2\pi}{p}\right)^2}.
\end{equation}
In this case, positive ($n>0$) and negative ($n<0$) diffraction orders give rise to two distinct resonance branches: one shifting toward shorter wavelengths and the other toward longer wavelengths, as shown in Fig.~3 (left). Moreover, because $\sin\theta<1$ the term proportional to $\sin\theta$ changes faster than the term proportional to $\sin\theta^2$, the resonance shift is more pronounced in the $\varphi = 0^{\circ}$ configuration than in the $\varphi = 90^{\circ}$ case.

The same behavior is observed in an experiment Fig.~5 (g,h).

\subsection*{Modeling}

The numerical results presented in the main text (Figs.~1--3) were obtained using the Fourier modal method implemented in the scattering-matrix formalism. In this section, the same approach is employed to reproduce the experimental results shown in Figs.~4 and 5.

The geometrical parameters of the grating used in the modeling were chosen close to those extracted from the SEM image in Fig.~4(a): the period is $p = 644$~nm, and the stripe widths are $a = 154.5$~nm and $b = 489.5$~nm. The thickness of the grating layer is $h_{\mathrm{gr}} = 290$~nm, and the thickness of the SiO$_2$ sublayer is $h_{\mathrm{SiO_2}}^{\mathrm{sub}} = 110$~nm. The Bragg mirror consists of alternating Si and SiO$_2$ layers with thicknesses, starting from the top Si layer $h^{\mathrm{DBR}} = [97,\ 260,\ 97,\ 260,\ 102]~\mathrm{nm}$.

In the experimental reflectivity spectra, the waveguide resonances appear broadened compared to the ideal numerical predictions. This broadening is attributed to fabrication-induced variations of the geometrical parameters within the illumination spot, as well as to material imperfections. To account for these effects in the simulations, a small effective absorption was introduced into the dielectric permittivities of the materials: $\varepsilon''_{\mathrm{Si}} = 0.02 i, \qquad
\varepsilon''_{\mathrm{SiO_2}} = 0.02 i$.

In the experimental data shown in Fig.~4(b,c), the resonances observed under nominally normal incidence exhibit a slight splitting. For an ideal alignment at $\theta = 0^{\circ}$ and $\varphi = 0^{\circ}$, such splitting is absent. Therefore, to reproduce the experimental spectra, we assumed a small deviation from normal incidence and performed simulations with an incidence angle of $\theta = 0.2^{\circ}$. The resulting spectra, shown in Fig.~\ref{fig:mod}(a,b), are in good agreement with the measurements.

For the configuration in which the plane of incidence is perpendicular to the stripes, the best agreement between experiment and modeling was obtained for an incidence angle of $\theta = 2^{\circ}$ and a small in-plane rotation angle of $\varphi = -2^{\circ}$. In the case where the plane of incidence is parallel to the stripes, a slight in-plane misalignment was also assumed, and the simulations were performed with $\varphi = 88^{\circ}$ instead of the ideal $\varphi = 90^{\circ}$. The corresponding results for an incidence angle of $\theta = 5^{\circ}$ are shown in Fig.~\ref{fig:mod}(e,f).

The remaining discrepancies between the experimental and simulated spectra can be attributed to several factors, including small uncertainties in the incidence angles during the measurements, deviations of the fabricated structure from the modeled geometry, spatial inhomogeneity of the fabricated sample, and uncertainties in the dielectric permittivities of the deposited materials, which are known to depend sensitively on the specific fabrication conditions.

\subsection*{Field distribution}

Figure~\ref{fig:fields} shows the electric-field intensity distributions $|E|^2$ in the $x$–$z$ cross section for $x$-polarized (top row) and $y$-polarized (bottom row) incident light. Each column corresponds to a different wavelength, illustrating the evolution of the field distribution inside and outside the operational band of the handedness-preserving mirror.

At a wavelength of $\lambda = 1200$~nm (bottom row, first column), a pronounced coupling of the incident field into the Si layers of the Bragg mirror is clearly observed. This behavior confirms the excitation of waveguide modes through grating-assisted diffraction, as discussed in the main text, and explains the appearance of sharp resonances that limit the short-wavelength edge of the operational band.

Within the handedness-preservation band, the field distributions for the two orthogonal linear polarizations exhibit markedly different spatial localization. The $x$-polarized field propagates predominantly within the air grooves of the grating, while the $y$-polarized field is strongly confined inside the high-index Si stripes. This polarization-dependent field localization introduces pronounced in-plane anisotropy, which results in a relative phase shift between the orthogonal polarization components upon reflection and enables the preservation of the circular polarization handedness.

At longer wavelengths (last column), the confinement of the $y$-polarized field within the Si stripes becomes weaker, and the field extends more uniformly across the structure. As a consequence, the effective anisotropy of the grating decreases, leading to a gradual degradation of the handedness-preserving performance at the long-wavelength edge of the operational band.

%\subsection*{Experimental setup}

\subsection*{Quarter-wave plate based handedness preserving mirror}
We theoretically considered a handedness preserving mirror consisting of a quarter-wave plate with a metallic mirror polished on one of the edges. The "Thorlabs' Superachromatic Wave Plates" have an extremely broad band from 600~nm to 2700~nm. Data from experimentally measured transmission, reflection, and theoretically calculated retardance at normal incidence and for oblique incidences is available on the manufacturer's site. For the second part of the system, we need a mirror that works in all entire band of the wave plate operation. The metallic mirrors have a very broad band and demonstrate reflection up to 97\%, however, only the silver mirror is an appropriate choice because the gold mirrors work up to 800 nm and aluminum mirrors have smaller reflectivity. For simplicity, we postulated the reflectance $R_{mirror} = 97\%$ at all wavelengths. 

The sum of reflection and transmission for a quarter-wave plate is not equal to unity, so we define the absorption of the plate as $A_{WP}(\lambda) = 1-T_{WP}(\lambda)-R_{WP}(\lambda)$. In the full structure under normal incidence, the light propagates inside the plate, reflects from the mirror, and travels back. The reflection of the system in that case equals the transmission of the plate with doubled absorption and doubled phase shift $R_{HPM} = T_{WP}\cdot A_{WP}\cdot R_{mirror}$. Fig.~\ref{fig:com}(a),(b) shows the reflection and retardance of the obtained handedness preserving mirror in all bands of the quarter-wave plate operation. As one can see in Fig.~\ref{fig:com} (c),(d), the theoretical reflection of our design (dashed line) is much higher than the experimental results for this hybrid structure (solid line). Moreover, the experimental results for our structure also give a reflectivity around 80\%.

Fig.~\ref{fig:com} (e) and (f) show the spectra of the theoretically calculated retardance for different angles of incidence. The solid lines are the results for a quarter-wave plate based mirror, and the dashed lines are the results calculated for our grating based structure. In panel (e), the plane of incidence is perpendicular to the stripes in our design, or perpendicular to the fast axis in a commercial anisotropic plate based mirror. In panel (f), the plate is aligned with the stripes or the fast axis. In a quarter-wave plate based mirror, the results strongly depend on the incidence angle and even for 10$^{\circ}$ decline, the reactance changes at 10\%. In contrast, our design for all angles results are almost coincident.
%%%%%%%%%%%%%%%%%%%%%%%%%%%%%%%%%%%%%%%%%%%%%%%%%%%%%%%%%%%%%%%%%%%%%%%%%%%%%%%%%%%%%%%%%%%%%%%%%%%%%%%%%%%%%%%%

\begin{figure*}[t!]
\centering\includegraphics[width=0.7\linewidth]{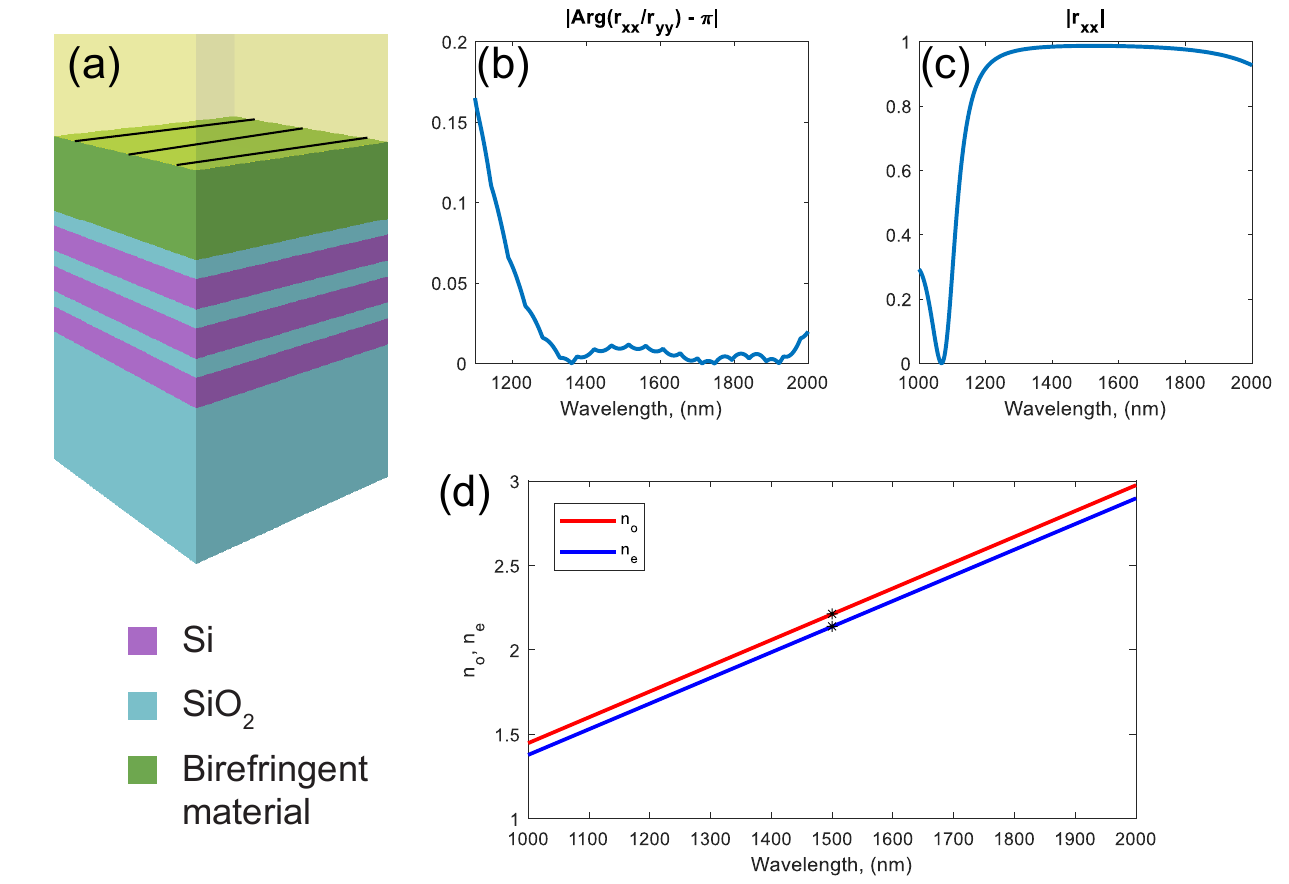}    
\caption{Handedness preserving mirror based on an anisotropic plate on Bragg mirror. (a) Scheme of the structure (black lines show the axis os anisotropy): thickness of the anisotropic layer $h_{\text{ani}} = 3810$~nm, the SiO$_{2}$ layers $h_{\text{SiO}_2} = 260$~nm and the Si layers $h_{\text{Si}} = 108$~nm. Spectral dependencies of (b) difference between $\arg(r_{xx}/r_{yy})$ and $\pi$ and (c) absolute value of $r_{xx}$. (d) Wavelength dispersion of refractive indexes for ordinary and extraordinary waves.}
\label{fig:anisplate}
\end{figure*}

\begin{figure*}[th]
\centering\includegraphics[width=1\linewidth]{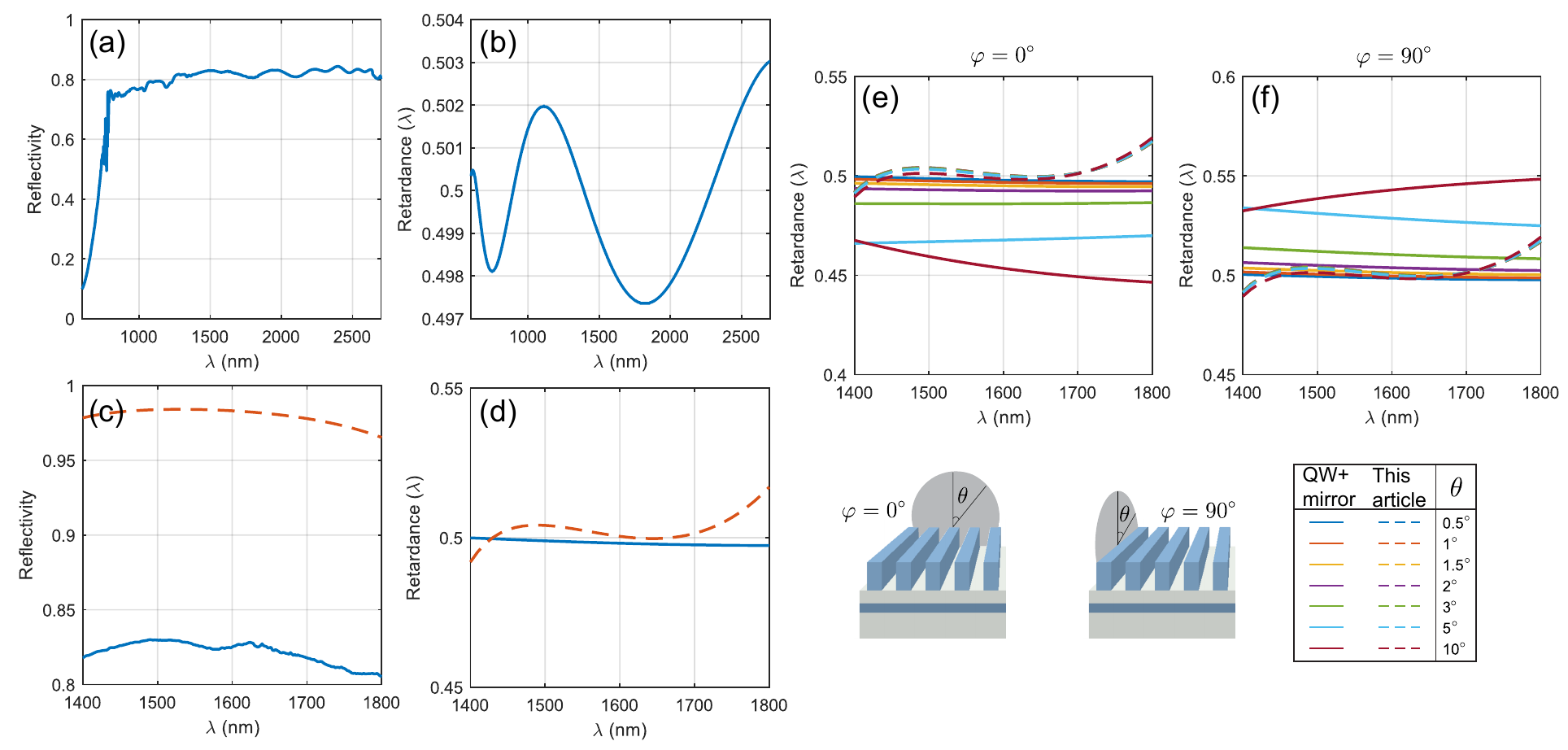}    
\caption{The spectral dependence of the (a) reflectivity and (b) retardance of the handedness preserving mirror consisted of commercial quarter-wave plate and mirror with reflectivity 98\%. the results are based on the experimental measurements of plate transmission available in the documentations. Comparison of (c) reflectivity and (d) retardance of mirror from (a),(b) and mirror made by our design in band of work for both mirrors. Comparison the spectral dependence of retardance of two mirrors for different angles of incident: (e) incidence perpendicular to the stripes in our design (dashed lines) and rotation perpendicular to fast axis in commercial plate based mirror (solid lines) and (f) incidence parallel to the stripes (dashed lines) and rotation about the fast axis (solid lines) respectively.}
\label{fig:com}
\end{figure*}

\begin{figure*}[t!]
\centering\includegraphics[width=0.8\linewidth]{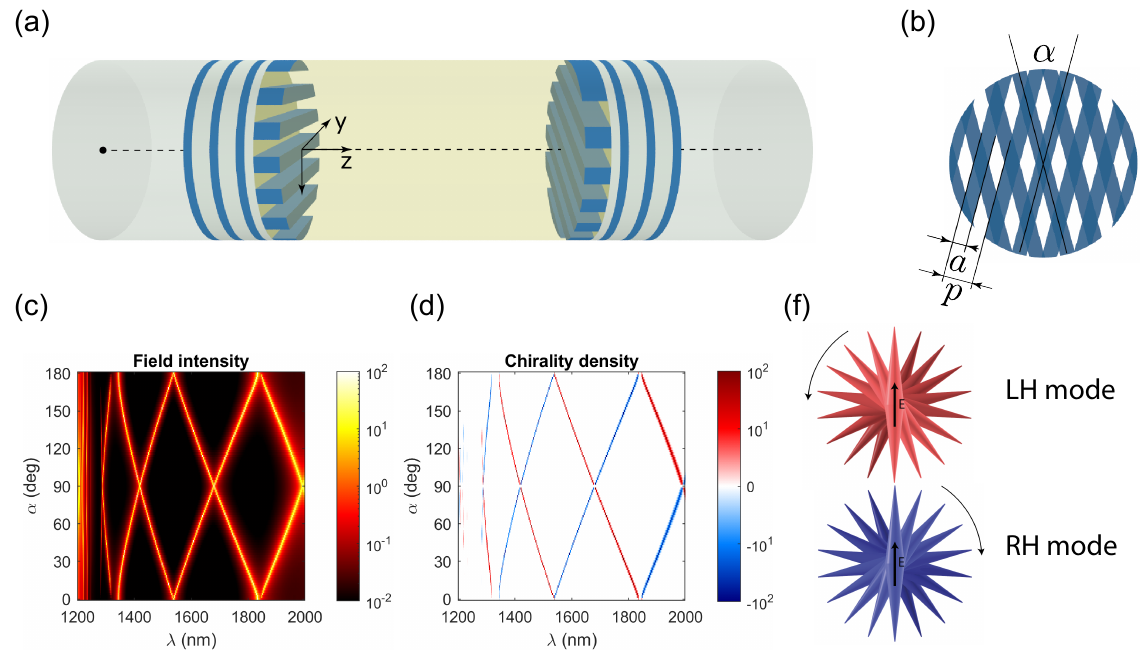}    
\caption{(a) Chiral Fabry-Pérot resonator consisted of two handedness preserving mirrors. (b) Two twisted gratings if look along $z$-axis. (c) Field intensity $I = (\vec{E}\cdot\vec{E}^*)+(\vec{H}\cdot\vec{H}^*)$and (d) chirality density $C = \text{Im}(\vec{E}\cdot \vec{H}^{*})$ calculated in the middle of the resonator. (f) Rotation of electric field vector inside the resonator for LH and RH modes if look along $z$-axis.}
\label{fig:FP}
\end{figure*}

%\bibliography{biblio}

\end{document}